\documentclass[aps,showpacs]{revtex4}
\usepackage{epsfig}

\begin{document}

\title{Exploring DNA translocation through a nanopore via a multiscale 
Lattice-Boltzmann Molecular-Dynamics methodology}
\author{Simone Melchionna$^1$, Maria Fyta$^2$, 
Efthimios Kaxiras$^{2}$, and Sauro Succi$^3$}
\affiliation{$^1$ INFM-SOFT, Department of Physics, Universit\`a di Roma 
\it{La Sapienza}, P.le A. Moro 2, 00185 Rome, Italy \\
$^2$ Department of Physics and Division of Engineering 
and Applied Sciences, Harvard University, Cambridge, MA, USA\\
$^3$ Istituto Applicazioni Calcolo, CNR, 
Viale del Policlinico 137, 00161, Roma, Italy
}

\begin{abstract}
A multiscale approach is used to simulate the translocation of DNA through 
a nanopore. Within this scheme, the interactions of the molecule with 
the surrounding fluid (solvent) are explicitly taken into account. 
By generating polymers of various initial configurations and lengths we map the
probability distibutions of the passage times of the DNA through the nanopore. 
A scaling law behavior 
for the most probable of these times with respect to length
is derived, and shown to exhibit an exponent that is in a good agreement 
with the experimental findings. The essential features of the
DNA dynamics as it passes through the pore are explored.

\end{abstract}


\maketitle

\section{Introduction}

An important biological process that has attracted the attention
of recent experimental studies is the translocation of a 
biopolymer such as DNA through narrow pores. These kind
of processes are not only important in phenomena like 
viral infection by phages, inter-bacterial DNA transduction or 
gene therapy,\cite{TRANSL}
but are also believed to open a way for ultrafast DNA-sequencing 
by reading the base sequence as the biopolymer passes through a nanopore.
Experimentally, it is possible to explore the translocation process 
through micro-fabricated channels under the effect of an external electric
 field, or through protein channels across cellular membranes.\cite{EXPRM}
Some universal features of DNA translocation can be analyzed by
means of suitably simplified statistical schemes \cite{statisTrans}
and non-hydrodynamic coarse-grained or microscopic models.\cite{DynamPRL,Nelson} 
However, a quantitative description 
of this complex phenomenon calls for state-of-the-art modeling.
Here, we describe a multiscale approach to 
capture the essential features of the generic problem of
a polymer translocating through a nanometer sized pore by taking
into account the effect of a surrounding solvent.
We are mainly interested in interpreting our results in relation
to experiments of DNA translocation through solid-state pores, which
are directed towards technological applications as was previously mentioned.
We also explore to what extent the results 
obtained describe the actual biological behavior of DNA translocation.

\section{Multiscale methodology and set-up}

In the present work a multiscale approach \cite{ourLBM} is used to 
trace the dynamic evolution of a polymer that is pulled through a narrow
pore in the presence of a fluid solvent. Within this framework,
methods involving different levels
of the statistical description of matter (for instance, continuum and 
atomistic) combine into a composite computational scheme.
The basic elements of the methodology used
are a Lattice Boltzmann treatment of the fluid solvent,\cite{LBE} and
a Molecular Dynamics simulation of the solute polymer. 
The solute-solvent dynamics are treated by coupling the
polymer to the hydrodynamic field of the surrounding solvent.
Owing to the dual field-particle nature of the Lattice Boltzmann 
technique, this coupling proceeds seamlessy in time and only requires 
standard interpolation/extrapolation for information-transfer 
in physical space. The scheme is general and applicable to any 
situation where a long polymer is moving in a solvent. 
This motion is of great interest for a fundamental
understanding of polymer dynamics in the presence of the solvent.
Previous studies dealt with hybrid models for polymer 
dynamics.\cite{hybrid1,hybrid2}
However, in these methods the coupling of the polymer
dynamics to the fluid solvent and the different time scales are not handled 
explicitly and as efficiently as in the current model. 
 These claims of efficiency are supported by the linear
scaling of the computational time with the number of polymer
beads and solvent degrees of freedom (see Ref.6).

\begin{figure}
\begin{center}
\includegraphics[width=0.5\textwidth]{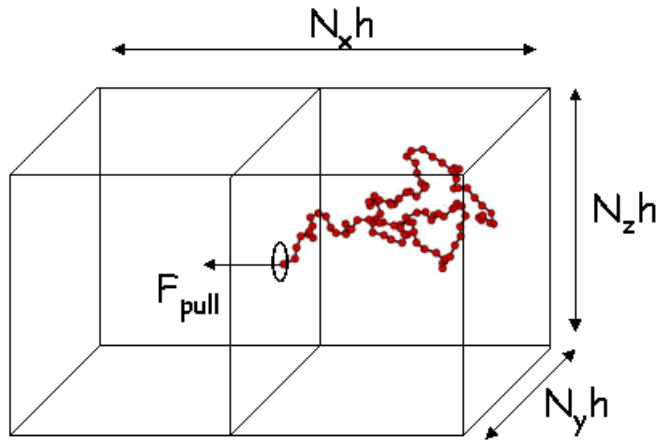}
\caption{\label{FIG1} Simulation set-up: The three dimensional box 
including the fluid and the polymer, as well as the wall that
separates the box into two chambers.
A random configuration of the polymer is shown at zero time, at which the
first bead is placed at the vicinity of the pore.
Translocation is induced by means of a pulling force $F_{pull}$.
Note that the box and the polymer are not in scale.}
\end{center}
\end{figure}

For our simulations, we consider a three-dimensional box of size
$N_x h \times N_y h \times N_z h$ 
lattice units, with $h$ the spacing between lattice points.
We choose $N_x = 2 N_y$, $N_y = N_z$, with $N_x=80$ or $100$
depending on the size of the translocating polymer chain.
The separating wall is located in the mid-section of the $x$ direction,
 at $x=h N_x/2$, and the polymer is initialized via a 
standard self-avoiding random walk algorithm and further relaxed 
to equilibrium by Molecular Dynamics. The solvent is 
initialized with the equilibrium distribution corresponding to a
 constant density and zero macroscopic speed. Periodic boundary
conditions are imposed for the fluid in all directions.
The polymer has a length in the range of $20 \leq N \leq 500$ beads, and
at $t=0$ resides entirely in the right chamber ($x>h N_x/2$).
The interaction of the polymer with the wall, as well as the 
interactions between non-adjacent beads
are modelled through Lennard-Jones potentials.
The bond length $b$ between consecutive beads is kept fixed at $1.2$.
 At the center of the separating wall, a square
hole of side $3h$ is opened, through which the
polymer can move from one chamber to the other.
Translocation is induced by a constant local electric field which
acts along the $x$ direction, and is confined in a rectangular
channel of size $3h \times h \times h$ along the streamline 
($x$ direction) and cross-flow ($y,z$ directions).
The three dimensional box considered in the simulations
is presented in Fig.\ref{FIG1}. A polymer configuration as well as the
separating wall are also shown. The direction of the translocation
is denoted by the arrow of the pulling force $F_{pull}$.
 The process falls in the fast translocation regime, where the total
 translocation time $t_X$ is much smaller than the typical Zimm time
of the polymer towards its minimum energy/maximum entropy
configuration. More details on the main parameters of the simulation 
are given elsewhere.\cite{ourLBM} Throughout this work, length and time 
are measured in units of the lattice spacing $h = \Delta x$ and 
time-step $\Delta t$.

\section{Passage times and statistics}
 
\begin{figure}
\begin{center}
\includegraphics[width=0.55\textwidth]{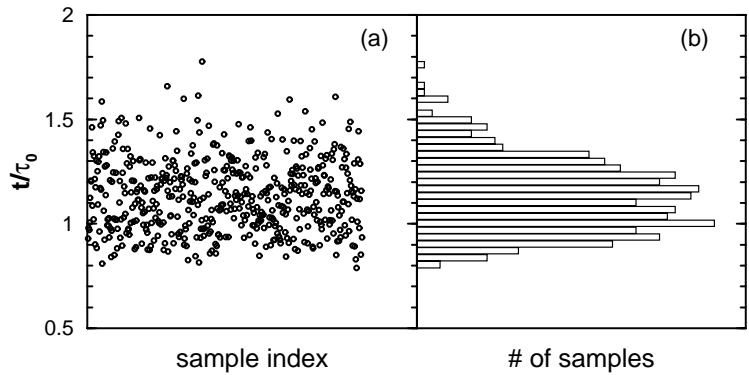}
\caption{\label{FIG2}(a) Translocation events and (b) projected probability 
distributions of the passage times for a polymer with N=100 beads.
Time is scaled with respect to the most probable time $\tau_0$.}
\end{center}
\end{figure}

We first turn to the derivation of the scaling behavior of the
translocation process. Simulations of a large number of
 translocation events ($100$ up to $1000$) for each polymer length
 were carried out. The ensemble of simulations is generated by 
different realizations of the initial polymer configuration
 and the variety of such events as well as their duration 
for $N=100$ beads
are shown in Fig.\ref{FIG2}(a). By accumulating all these events 
for all lengths, duration histograms were constructed. These
are not gaussians but are rather skewed towards longer times. 
This is clearly seen in Fig.~\ref{FIG2}(b), where the events 
in Fig.~\ref{FIG2}(a) have been projected into a duration
histogram. Accordingly, we use the most probable time (peak of the
distribution) as the representative translocation time for 
every distribution; this is also the definition of the 
translocation time in experiments, to which we compare later.
Calculating the most probable times for each length
leads to a nonlinear relation between the
translocation time $\tau_0$ and the number of beads $N$:
$\tau_0 (N) \propto N^{\alpha}$, with an exponent $\alpha \sim 1.29$.
Preliminary results of simulations in the absence of a solvent give an 
exponent over $1.4$, similar to theoretical work based on Brownian
 models. \cite{KARDAR} Such a difference indicates a noticeable
acceleration ot the process due to hydrodynamic interactions.
Overall, our results are quite similar to the corresponding experimental
 data for DNA translocation through a nanopore,\cite{NANO} which we 
will discuss in the last section.

\section{Polymer Dynamics}

Additional insight can be gained by analyzing the polymer dynamics
 during translocation. 
The molecule shows a blob-like conformation as it moves
through the hole. It may either translocate very fast or move from 
one chamber to the other intermittently, with pauses.
The fluid imposes fluctuations to the polymer motion and these 
are correlated to the entropic forces (gradient
of the free energy with respect to a conformational order parameter, 
typically the fraction of translocated beads.
These forces act on both translocated and untranslocated parts of the polymer. 

\begin{figure}[h!]
 \begin{center}
\includegraphics[width=0.65\textwidth]{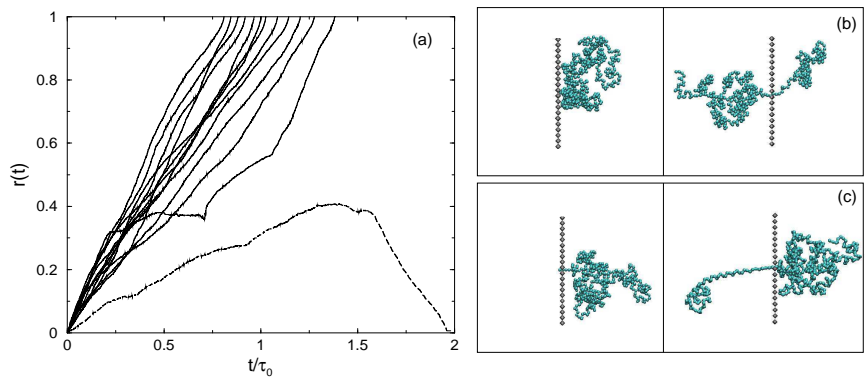}
 \caption{\label{FIG3}Progress in time of $r(t)$ (see text)
 for $N=500$ monomers. (a) Curves corresponding to translocation events
for various initial configurations. (b) A typical configuration
that translocates successfully, shown at about $r(t)$=0.75 (right).
On the left the initial configuration ($r(t)$=0) of the same polymer is shown.
Time is scaled with respect to the most probable time $\tau_0$ for
these events. (c) A retraction event is shown for a different set
of parameters. The configuration for this event is shown at a timestep 
where it has already started to reverse its motion (right).
The initial configuration of this retracted polymer
is sketched on the left.}
 \end{center}
 \end{figure}

It is very instructive to monitor the progress in time of the
number of translocated monomers $N(t)$: $r(t) \equiv N(t)/N$
serves as a reaction coordinate, with the total translocation time 
defined by the condition $r(t_X)=1$. The translocated monomers
are plotted in Fig.\ref{FIG3}. The events described by different
$r(t)$ curves correspond to various initial configurations
and represent completed translocation events.
The translocation for a given polymer proceeds along a curve
virtually related to its initial configuration and its interactions with
the fluid. A configuration for a typical event at a specific 
timestep is shown in Fig.\ref{FIG3}(b). The initial
conformation of the same polymer is shown in the left of this panel.
In the same figure $r(t)$ for a retraction event is represented by 
a curve which returns to 0.
In this case, some of the parameters in the simulations 
have changed: 
specifically, the temperature was decreased to
kT=10$^{-5}$ instead of 10$^{-4}$, the Lennard-Jones coupling was stronger
(0.002 instead of 10$^{-4}$) and the pulling force was reduced 
from $0.02$ to $0.01$. 
All numbers are given in simulation units.\cite{ourLBM}
The parameter space largely remains to be further explored, and this
exploration is expected to lead to a host of new interesting features
in the dynamics of the translocating polymer.
The configuration for the retracted case is also shown (Fig.\ref{FIG3}(c))
at a time where it already has started to reverse its motion. 
The initial configuration for this polymer is presented in the
left of the same panel.
An elongated conformation is clearly visible, that delays the translocation, 
leading to an increase of the entropic force acting at
the untranslocated part of the chain which eventually leads
to polymer retraction.
However, inspection of all initial polymer conformations
as the ones depicted in Fig.\ref{FIG3} cannot lead to a classification of
these configurations and a subsequent prediction of 
the corresponding $r(t)$ curves.

\begin{figure}
\begin{center}
\includegraphics[width=0.7\textwidth]{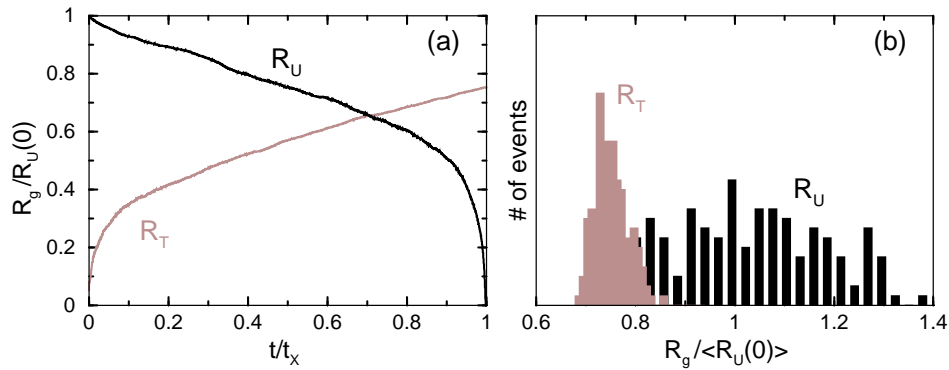}
\caption{\label{FIG4}Translocated ($R_{T}$) and untranslocated ($R_{U}$) 
radii of gyration. (a) A typical translocation event. 
(b) The distribution for all the events for $N=400$.
Time and radii are scaled with respect to $t_X$ and $R_{U}(t=0)$, respectively.
$R_U$ and $R_T$ are scaled with respect to the mean value of $R_{U}(t=0)$.}
\end{center}
\end{figure}

The dynamics of the translocating polymer can be further examined by
analyzing the radius of gyration. To this end, in 
Fig.\ref{FIG4}(a) we represent
 the time-evolution of the radii of gyration $R_g$ for the two sections 
of the polymer, the untranslocated ($R_U(t)$) [$x>h N_x/2$] and translocated 
$R_T(t)$ [$x<h N_x/2$], respectively. 
The curves related to these two sections differ considerably. 
By definition, $R_U(t)$ vanishes at $t_X$, but at the same time
$R_T$ increases, although not up to the value $R_U(t=0)$. 
This feature indicates a very different behavior of polymer at the 
beginning and at the end of the translocation process. 
The polymer retains its blob-like configuration, but its final
volume is smaller than it was initially, at $t=0$.
This is the result of the strong entropic constraint set by the presence
of the hole. The distribution of the radii of gyration for 
all the events is shown in Fig.\ref{FIG4}(b), 
for both the untranslocated part at zero
time, $R_U(t=0)$ and the translocated one at the end of the
process, $R_T(t=t_X)$. This histogram clearly shows that at the
end of the process the polymer becomes more confined than it
was initially at $t=0$. A detailed explanation would require
taking into account the many-body correlations and the non trivial 
interplay between all forces acting on the polymer.
It is conceivable that, by allowing the polymer to further advance
in time, it will regain its initial volume, but this remains to be examined.
Nevertheless, in this work we focus on the first-passage times,
namely up to the time needed for the last bead to translocate through the pore.

\section{Connection to experiment}

As was already mentioned, our results are quite similar to the
experimental data for DNA translocation through a nanopore.\cite{NANO}
An encouraging comparison is that the scaling we found 
for the translocation time with polymer length (with $\tau_0\propto N^\alpha$,
$\alpha = 1.29$) is quite close to the experimental measurement
for DNA translocation ($\alpha = 1.27 \pm 0.03$).\cite{NANO}
Previous studies used different interpretations of their
 coarse-grained simulations of DNA packing in
 bacteriophages \cite{beadParam1} or $\lambda$-phage DNA 
 in solution \cite{beadParam2}: one bead can 
 typically represent a number of base-pairs (bp), ranging
 from about 8 base-pairs (similar
 to the hydrated diameter of B-DNA in physiological
 conditions) to $\sim10^3$ base-pairs. 
 
In order to interpret out results in terms of physical units,
we calculated the persistence length ($l_p$) of the semiflexible 
polymers used in our simulations. Accordingly, the formula 
for the fixed-bond-angle model of a worm-like chain was 
used \cite{lpWCL}: $l_p=b/(1-\cos\langle \theta \rangle)$, 
where $\langle\theta\rangle$ is complementary to the average 
bond angle between adjacent bonds. In lattice units 
($\Delta x$) an average persistence length for the polymers
considered, was found to be approximately $12$.
For $\lambda$-phage DNA $l_p\sim 50$ nm \cite{lpDNA}.
This is equated to $l_p$ for our polymers, thus the 
lattice spacing is $\Delta x \sim 4$ nm, which is
also the size of one bead. Thereby, one bead maps
approximately $12$ bp, given that the base-pair spacing is
$0.34$nm. With this mapping, the pore size is
about $\sim~12$ nm, close to the experimental pores
which are of order $10$ nm. The polymers presented here correspond to
DNA lengths in the range $0.2-6$ kbp. The DNA lengths used
in the experiments are larger (up to $\sim100$ kbp);
the current multiscale approach can be extendet
to handle these lengths, assuming that appropriate
computational resources are available.
Additional details on the mapping of the anonymous polymers
simulated in this work to real biopolymers will be given elsewhere.

\section{Conclusions}

A multiscale methodology has been applied to study polymer translocation
through a narrow pore in the presence of a fluid solvent.
Special emphasis has been placed on the passage times and 
their scaling with the polymer length. 
Hydrodynamic interactions assist translocation, and the dynamics
of the polymer during the process was monitored.
Further exploration of the parameter space of our model reveals interesting
features, like the retraction of the polymer chain.
Visual inspection of the detailed dynamics of the simulations suggests a
rather complex picture for the atomic-scale mechanisms
underlying the translocation process.
Comparison of our results revealed encouraging similarities with
experiment on DNA translocation through narrow pores, especially in 
the scaling law of the translocation time with polymer length.
A thorough exploration of parameter space is required in order to 
permit direct contact with experimental results. Nevertheless, our 
results so far are encouraging in that a multiscale approach seems 
capable of providing a realistic description of DNA translocation 
through a nanopore and promises to yield interesting insight into
this complex process.

\section*{Acknowledgments}
M.F. acknowledges support by the Nanoscale Science
and Engineering Center, funded by the National Science Foundation
under Award Number PHY-0117795.

\end{document}